# Resolving and routing the magnetic polymorphs in 2D layered antiferromagnet

Zeyuan Sun[1†], Canyu Hong[1†], Yi Chen[2], Zhiyuan Sheng[1], Shuang Wu[1], Zhanshan Wang[1], Bokai Liang[1], Wei-Tao Liu[1], Zhe Yuan[3], Yizheng Wu[1, 4], Qixi Mi[2], Zhongkai Liu[2, 5], Jian Shen[1, 3, 4], Shiwei Wu[1, 3, 4*]

[1] State Key Laboratory of Surface Physics, Key Laboratory of Micro and Nano Photonic Structures (MOE), and Department of Physics, Fudan University, Shanghai 200433, China
[2] School of Physical Science and Technology, ShanghaiTech University, Shanghai 201210, China
[3] Institute for Nanoelectronic Devices and Quantum Computing, and Zhangjiang Fudan International Innovation Center, Fudan University, Shanghai 200433, China
[4] Shanghai Research Center for Quantum Sciences, Shanghai 201315, China
[5] ShanghaiTech Laboratory for Topological Physics, ShanghaiTech University, Shanghai 200031, China

[†] These authors equally contributed to this work.
[*] Corresponding email: swwu@fudan.edu.cn

**Abstract:** Polymorphism, commonly denoting the variety of molecular or crystal structures, is a vital element in many natural science disciplines. In van der Waals layered antiferromagnets, a new type of magnetic polymorphism is allowed by having multiple layer-selective magnetic structures with the same total magnetization. However, resolving and manipulating such magnetic polymorphs remain a great challenge. Here we use the phase-resolved magnetic second-harmonic generation microscopy to elucidate such magnetic polymorphism in the 2D semiconducting layered antiferromagnet chromium sulfur bromide (CrSBr), and demonstrate how the magnetic polymorphs can be deterministically switched in an unprecedented layer-selective manner. With the nonlinear magneto-optical technique unveiling the magnetic symmetry information through the amplitude and phase of light, we could unambiguously resolve the polymorphic spin-flip transitions in CrSBr bilayers and tetralayers. Remarkably, the deterministic routing of polymorphic transitions originates from the breaking of energy degeneracy via a magnetic "layer-sharing" effect: the spin-flip transitions in a tetralayer are governed by the laterally extended bilayer, which acts as a "control bit". We envision such controllable magnetic polymorphism to be ubiquitous for van der Waals layered antiferromagnets, and could lead to conceptually new design and construction of spintronic and opto-spintronic devices for probabilistic computation and neuromorphic engineering.

The discovery of two-dimensional (2D) layered antiferromagnets provides an exciting platform, allowing the manipulation of magnetic states layer-by-layer[1-6]. In these materials, each monolayer behaves as an Ising ferromagnet with dual magnetic states (e.g. '←' and '→'), which can act as a binary bit. This ferromagnetic (FM) ordering is primarily driven by intralayer super-exchange interaction, and the magnetic anisotropy arises from the spin-orbit coupling[7-9]. As more layers are stacked, the interlayer antiferromagnetic (AFM) coupling leads to the ground state with antiparallel magnetic alignment across the van der Waals (vdW) interfaces (Fig. 1a). This AFM coupling strength is typically comparable in magnitude to the intralayer magnetic anisotropy[6]. The subtle balance between these interactions is therefore crucial in determining the overall magnetic states, providing the opportunity to control the magnetization in each layer via external stimuli such as magnetic field[1-3,10-12], electric field[13-16], stacking order[2,17,18] and strain engineering[19-22].

Among these 2D layered antiferromagnets, CrSBr emerges as a representative material and attracts great attentions owing to its semiconducting behavior and magneto-excitonic states[4,22-29]. In monolayer CrSBr, the strong intralayer FM aligns along the *b*-axis (Fig. 1b). Conceptualizing each monolayer as a binary magnetization $M = \pm 1$, the total number of magnetic states could increase exponentially with additional number of layers, which also gives rise to magnetic polymorphism with degenerate states. For example, at $M = 0$, a bilayer can have two distinct AFM states (Fig. 1c), while a tetralayer can have six (Fig. 1d). Interestingly, the enumeration of these magnetic polymorphs follows the mathematical concept of *Yang Hui*'s Triangle (also known as *Pascal*'s Triangle), as depicted in Fig. 1e. The number of layers *n* and total magnetization *M* become the basis for determining the total number of magnetic polymorphs. This combinatorial nature may unlock many promising spintronic applications in both fundamental science and innovative technologies, including applied mathematics, advanced computing and artificial intelligence[30,31].

Despite huge prospects, the layer-resolved identification of magnetic polymorphs is highly challenging, as conventional experimental approaches hit obstacles in accurately characterizing the magnetic polymorphs. Techniques such as neutron scattering are not suitable for detecting 2D magnets down to atomic thickness and micron lateral size. Linear magneto-optical methods like the magneto-optical Kerr effect (MOKE) and magnetic circular dichroism (MCD) are sensitive to non-zero net magnetization, but yield negligible or weak signal on antiferromagnets[1]. Similar problems exist for various scanning probe microscopic techniques such as magnetic force microscopy[32] and NV center magnetometry[33,34]. Although photoluminescence (PL), reflectance spectroscopy and recent tunneling magnetoresistance measurement can effectively sense the change in magnetization, they fall short in distinguishing the magnetic polymorphs that have degenerate electronic structures[23,29,35-37].

Optical second-harmonic generation (SHG) has recently emerged as a powerful tool to study the magnetic symmetry and property in layered antiferromagnets[17,38-41]. As a coherent optical process, both the amplitude and phase of SHG are essential in retrieving the spatial and temporal symmetry properties of the medium. Unfortunately, the phase information is often lost, as most experiments detect the SHG intensity only. In this study, we apply the phase-resolved SHG (phase-SHG) technique to determine the magnetic structures of few-layer CrSBr, in an unprecedented layer-selective manner. In particular, we successfully identify the magnetic polymorphs in CrSBr bilayer and tetralayer with microscopic imaging and phase-shifting interferometry. Moreover, the magnetic polymorphs in isolated tetralayer samples are degenerate, as evidenced by the non-repetitive spin-flip transitions and stochastic domain formations during the magnetic field sweep along the easy axis. In sharp contrast, the tetralayer with a laterally extended bilayer shows highly repetitive spin-flip transitions, with two types of magnetic polymorphs emerging at $M = \pm 2$. Further investigation of such layer-adjacent samples highlights the key role of "layer-sharing" effect, in which the associated bilayer contributes a decisive intralayer FM coupling extending into the tetralayer. This bilayer not only routes the specific magnetic polymorph of the associated tetralayer, but also anchors the tetralayer magnetization that extends for tens of micrometers within the 2D plane.

**Resolving the magnetic polymorphs**

The CrSBr few-layer samples are mechanically exfoliated onto the oxidized silicon wafer. The strong magneto-excitonic coupling enables the detection of spin-flip transitions through PL spectroscopy under external magnetic field[23,37]. Figure 2a shows the PL hysteresis loop of a CrSBr bilayer. The external magnetic field is always along the $b$-axis (see Fig. 1b) and the PL intensity is obtained by integrating over the wavelength range of excitonic resonance (Fig. 2b). The PL loop shows step-like switching behaviors, indicating sudden changes of electronic structure caused by spin-flip transition. At 0 T, the PL spectra in both the forward and backward sweeps exhibit a broad excitonic peak at 1.34 eV, while this peak red-shifts to 1.32 eV under ±0.6 T (Fig. 2b). The two identical spectra at 0 T indicate the same excitonic state. Nonetheless, it may arise from two degenerate AFM polymorphs that could not be resolved from PL spectra.

Due to the distinct magnetic symmetry of AFM and FM states, SHG serves as a unique method to characterize them in layered antiferromagnets[17,39]. In even-layer CrSBr, despite that its crystallographic structure is centrosymmetric[39], the AFM states break inversion symmetry and give rise to electric-dipole (ED) allowed SHG with time-noninvariant c-type second-order nonlinear susceptibilities $\chi_{ijk}^{ED}$. In sharp contrast, the FM states retain the inversion symmetry and yield negligible SHG signal. Figure 2c indeed shows the SHG loop of the bilayer with a sudden intensity drop upon reaching the critical fields, whose values are also consistent to those in the PL loop. By measuring the SHG excitation spectroscopy (Fig. 2d), the AFM states at 0 T exhibit a single peak when the fundamental photon energy is in resonance with the excitonic transition. However, in contrast to PL spectra, the SHG intensities at 0 T are clearly different between the forward and backward sweeps, suggesting that the bilayer may undergo different AFM states.

To understand the observation, we examine the symmetry of the two degenerate AFM states in CrSBr bilayer, which form a pair of time-reversal or spatial-inversion counterparts, or magnetic polymorphs. Between them, $\chi_{ijk}^{ED}$ has a sign change and thus a $\pi$ phase shift in the SHG responses (see Supplementary Information)[17,41-43]. This phase shift is often lost in intensity measurements, but can be revealed by the phase-SHG technique[44,45]. In our setup (Extended Data Fig. 1), the phase-SHG employs an interference between two SHG signals: one from the CrSBr sample and the other from a reference such as y-cut quartz. The SHG polarizations from CrSBr and quartz are orthogonally aligned along the fast and slow axis of a Soleil-Babinet compensator (SBC), which introduces a tunable external phase between the two respective signals from CrSBr and quartz. These two light fields are then superimposed by a polarization analyzer before optical detection. Through the modulation of external phase, intensity variations can be generated, allowing the observation of a phase shift between the two AFM states.

Figure 2e shows the phase-SHG hysteresis loop of the CrSBr bilayer. The loop exhibits the same switching fields as in the PL and SHG measurements, but with contrasting intensities at the two AFM states during the forward and backward sweeps. The non-zero intensities at both FM states are due to the reference quartz. The inset presents phase-SHG images at the two AFM states, displaying destructive and constructive interferences between SHG from the bilayer and reference quartz, respectively. Moreover, the phase-SHG intensities at AFM states change sinusoidally with the tuning of external phase via SBC (Fig. 2f), confirming a $\pi$ phase shift between the two time-reversal or spatial-inversion counterparts. Further adjustment of SBC for reversed contrast exhibits the same switching behavior (Extended Data Fig. 2). These results provide unambiguous evidence for the existence of two distinct AFM states – a pair of magnetic polymorphs. The deterministic switching between two polymorphs in the bilayer indicates the "soft" and "hard" layers with different magnetic anisotropy, which can be attributed to the asymmetric interfaces with vacuum and the substrate, respectively. This interfacial asymmetry is also responsible for the SHG intensity difference between the two antiferromagnetic polymorphs in Fig. 2d.

**Routing the layer-selective magnetic states**

Compared to bilayer, CrSBr tetralayer exhibits more complex magnetic behaviors with 16 permutations of layered magnetic structures, as listed in Extended Data Table 1. These permutations result in magnetizations of $M = 0$, $\pm 2$ and $\pm 4$, with no odd values due to the layer-by-layer spin-flip transitions. Based on the interlayer magnetic couplings, we classify the magnetic structures into four types. The AFM ground states with $M = 0$ contain three AFM vdW interfaces (FM# = 0, AFM# = 3), and the fully-aligned FM states with $M = \pm 4$ contain three FM interfaces (FM# = 3, AFM# = 0). For $M = \pm 2$, two types of structures emerge, with the 'Type-I' states having one FM and two AFM interfaces (FM# = 1, AFM# = 2), in comparison to 'Type-II' states with two FM and one AFM interfaces (FM# = 2, AFM# = 1). These two types have the same magnetization and form the magnetic polymorphs. Except for the FM states, the other three types of magnetic structures break the inversion symmetry, each with unique symmetry properties that can be resolved by SHG measurements.

We first study the SHG response in isolated CrSBr tetralayers (Fig. 3a). The samples are separated from other flakes of different thickness, and uniform in optical microscopic images (Fig. 3b). However, subsequent field sweeping cycles reveal stochastic domain formations in SHG images (Fig. 3c and Extended Data Fig. 3). Multiple SHG hysteresis loops at fixed sample position exhibit non-repetitive transition behaviors (Fig. 3d). Similar phenomena are also observed in the PL hysteresis loops (Extended Data Fig. 4). Despite the variations of switching field and SHG/PL intensity among different hysteresis loops, the CrSBr tetralayers undergo the same number of sudden switchings, indicating the layer-by-layer sequential spin-flip transitions. Nevertheless, all these observations suggest intricate magnetic couplings among different layers in isolated tetralayers.

In sharp contrast, the magnetic transition behaviors are repetitive in non-isolated CrSBr tetralayers. Figure 3e and f illustrate a non-isolated tetralayer with a laterally extended bilayer.

In this layer-adjacent sample, the tetralayer exhibits a uniform single domain during the magnetic transitions (Fig. 3g and Extended Data Fig. 5). Several SHG loops consistently present multiple step-like switchings at the same magnetic fields, corresponding to transitions between the magnetizations of $\boldsymbol{M} = 0, \pm2$ and $\pm4$ (Fig. 3h). Such repetitive transition behaviors in layer-adjacent sample suggest that the magnetic coupling in tetralayer may be influenced by the laterally extended bilayer. This bilayer breaks the degeneracy of magnetic polymorphs in tetralayer, resulting in deterministic switching between specific magnetic states. Further examining the SHG loops, the intensity differences at $\boldsymbol{M} = 2$ or -2 indicate the presence of both 'Type-I' and 'Type-II' magnetic structures at the corresponding magnetizations.

To gain further insight into the specifics of these magnetic states, we also acquire the PL spectra under different magnetic states. The switching fields in PL loop show a one-to-one correspondence to those in SHG loops (Fig. 4a). The PL spectra at AFM ground states exhibit a major peak at 1.34 eV (Fig. 4b), which shifts to 1.32 eV in FM states (Fig. 4c). For states with $\boldsymbol{M} = \pm2$, the spectra also display two distinct types and intensities (Fig. 4d,e). One of them resembles to that of the AFM states and the other approaches to that of FM states. We therefore assign the former to 'Type-I' states (FM# = 1, AFM# = 2) and the latter to 'Type-II' states (FM# = 2, AFM# = 1), because the magneto-excitonic transition in CrSBr sensitively depends on the interlayer magnetic coupling. The variations of magneto-excitonic behaviors between different types of magnetic states are also reflected in the SHG excitation spectra shown in Extended Data Fig. 6, which lead to the change of SHG intensities between 'Type-I' and 'Type-II' states (Fig. 3h).

While PL and SHG intensity measurements differentiate the 'Type-I' and 'Type-II' states at $\boldsymbol{M} = \pm2$, they still cannot fully resolve the specific magnetic structures. For a given magnetization, each type of states permits a pair of magnetic polymorphs that are transformed by spatial-inversion, as separated by dashed lines in Extended Data Table 1. To further resolve these magnetic polymorphs and investigate their deterministic switching in layer-adjacent samples, we perform the phase-SHG measurements. Figure 5a,b show the phase-SHG hysteresis loops in the bilayer and tetralayer. These loops distinguish not only the AFM ground states in bilayer, but also the AFM, 'Type-I' and 'Type-II' states in tetralayer. Notably, the switching fields of bilayer are slightly smaller than those of tetralayer from AFM to '$\boldsymbol{M} = \pm2$' states, suggesting sequential spin-flip transitions initiate from the bilayer and then follow by the tetralayer.

The phase-SHG interferograms for tetralayer in Fig. 5c-e show that the AFM, 'Type-I' and 'Type-II' states form three pairs of magnetic states, with each state exhibiting $\pi$ phase shift from its counterpart. These observations illustrate the time-reversal symmetry in each pair, and further help deduce the evolution of magnetic transitions. For clarity, the layers from bottom to top are labelled as $1^{st}$ to $4^{th}$, with the bottom two layers in tetralayer extending out to the bilayer (Fig. 5f). Considering each switching involves a single-layer spin-flip transition, the magnetic evolution starts with a spin-flip in the $3^{rd}$ layer of FM state (step #1), generating the 'Type-I' state. This is followed by a flip in the $1^{st}$ layer, which transitions both the bilayer and tetralayer into AFM ground states (step #2). Subsequently, a spin-flip in the $2^{nd}$ layer of the bilayer triggers a corresponding flip in the tetralayer (step #3), resulting in the 'Type-II' state. The final transition occurs when the $4^{th}$ layer flips, switching the tetralayer into the other FM state (step #4). A backward field sweep reverses these steps as #1'-#4', mirroring the time-reversal magnetic structures of step #1-#4.

The route of magnetic transitions satisfies the constraints depicted by the phase-SHG interferograms and facilitates a notable layer-sharing effect between the bilayer and tetralayer. This effect extends intralayer ferromagnetism laterally in layer-adjacent samples, resulting in deterministic spin-flip transitions. We note that an alternative magnetic evolution in Extended Data Fig. 7, consisting of spatial-inversion counterparts at $M_{4L} = 0, \pm2$, also fits the phase-SHG interferograms. However, this route is unlikely because the $1^{st}$ layer of tetralayer should have a lower spin-flip energy barrier than the $3^{rd}$ layer in the steps #3 and #3', and this route breaks the intralayer FM coupling between the bilayer and tetralayer.

**Discussion on the layer-sharing effect**

To rationalize the routing mechanism via the layer-sharing effect, we focus on the energetics involved in the deterministic spin-flip transitions, consisting of the intralayer FM coupling $J_\parallel$ and interlayer AFM exchange coupling $J_\perp$, respectively. During the transitions from FM to 'Type-I' states (steps #1 and #1'), the layer-sharing effect stabilizes the magnetizations in the $1^{st}$ and $2^{nd}$ layers because of the strong intralayer FM coupling $J_\parallel$ between the bilayer and tetralayer. On the other hand, the $3^{rd}$ layer flips its magnetization to favor the double interlayer AFM exchange couplings with the $2^{nd}$ and $4^{th}$ layers.

For the AFM to 'Type-II' transitions (steps #3 and #3'), the $2^{nd}$ layer in bilayer flips first when the applied magnetic field overcomes the AFM exchange coupling. The firstly-flipped $2^{nd}$ layer creates a magnetic domain wall at the bilayer-tetralayer boundary, separating two regions with opposite magnetizations. Considering the $2^{nd}$ layer as a ferromagnet, with the dominance of $J_\parallel$ over $J_\perp$, the total energy of this layer favors to expand the domain that is aligned with the magnetic field, and to diminish the domain that is against the field[46]. The energy threshold for this process is much lower than a simultaneous spin-flip of the entire domain. Therefore, a slightly exceeding field causes the domain wall to move into the tetralayer, leading to the magnetic transition of tetralayer. In this scenario, the layer-sharing effect enables a unified 'Type-II' domain to extend laterally up to 20 μm from the bilayer-tetralayer boundary (Extended Data Fig. 5). Here the bilayer effectively acts as a "control layer" for the initiation of 'Type-II' transitions in the tetralayer. More interestingly, if a tetralayer is laterally sandwiched by a bilayer on one side and a thicker layer on the other side, as exemplified in Extended Data Fig. 8, the domain wall propagation may hit an obstacle, forming multiple domains such as 'Type-I' and 'Type-II' in the tetralayer.

To further illustrate the layer-sharing effect, we apply the femtosecond laser cutting technique to a few-layer CrSBr and *in-situ* image the impact of adjacent layers on magnetic domains by magneto-PL. Figure 6a shows the optical microscopic image of the few-layer CrSBr, consisting of a tetralayer laterally connected to neighboring bilayer, trilayer and thicker layers. PL images on the tetralayer show the appearance of two types of magnetic domains at intermediate magnetic fields of -0.25 T (forward) and 0.25 T (backward), corresponding to the Type-I (lower PL intensity) and Type-II (higher PL intensity) states (Fig. 6b,c). After we apply the *in-situ* laser cutting along the lines shown in Fig. 6d, the magnetic domains on the tetralayer emerge into a single one (Fig. 6e,f). This set of experimental data clearly visualizes the dominant role of layer-sharing effect in determining the domain walls and magnetic polymorphs in CrSBr few-layers.

To summarize, our results highlight the magnetic polymorphism as an inherent phenomenon in vdW layered antiferromagnets. The distinctive phase detection of SHG thus offers an indispensable method for investigating intricate magnetic interactions among these magnetic polymorphs. Although the demonstrated number of states is only eight in CrSBr tetralayers, the layer-sharing effect can potentially manipulate numerous magnetic polymorphs in thicker layers via the local control of laterally extended bilayer or monolayer. The exploration of the layer-sharing effect and magnetic polymorphism, assisted by the laser cutting technique, could revolutionize the conventional construction of spintronic and opto-spintronic devices[47], providing a new strategy for the design of innovative applications such as probabilistic computation[48] and neuromorphic engineering[49].

## Methods

### Crystal growth

A quartz tube (Ø25 mm, 200 mm long) was charged with Cr powder (99.95%, 1.040 g, 20 mmol) and $S_2Br_2$ (2.349 g, 10.5 mmol, from S and $Br_2$). The Cr powder on one end of the tube was ignited by a butane flame to initiate the reaction with $S_2Br_2$. Once the reaction subsided, the tube was placed in a two-zone tube furnace and heated to 930 °C and 880 °C, respectively. After 3200 min, metallic black, blade-like crystals formed in the middle of the tube, distinct from other products ($Cr_2S_3$ and $CrBr_3$) at two ends. The CrSBr crystals were washed with ethanol and stored under ambient conditions. The identity of the sample was confirmed by single-crystal X-ray diffraction (Bruker D8 VENTURE, Mo $K\alpha$ radiation).

### Sample preparation

All the CrSBr samples were mechanically exfoliated onto $Si/SiO_2$ wafers inside a nitrogen-filled glovebox. The layer thickness was determined by the optical contrast and atomic force microscopy (Extended Data Fig. 9). To avoid the exposure to air during the sample mounting to the optical cryostat, the samples were placed inside a copper cave and sealed by borosilicate cover glass with vacuum grease. For the phase-SHG measurements, the cover glass was replaced by a 0.2 mm thick y-cut quartz, and the gap between quartz and wafer was finely adjusted to 70-150 μm by using an optical microscope inside the glovebox. The *in-situ* laser cutting was performed at base temperature (~6 K) and under high vacuum (<$1\times10^{-8}$ Torr). A femtosecond laser with a wavelength at 925 nm and a power of ~80 mW was used, focused through a 50× objective (NA = 0.55). Each cutting path was executed in a few seconds to minimize potential damage and thermal effects.

### PL and SHG measurements

The experiments were conducted in a home-built variable temperature magneto-optical cryostat under high vacuum (<$1\times10^{-8}$ Torr). The sample temperature was ~6 K unless otherwise mentioned. The cryostat was held inside a room-temperature bore superconducting magnet with 7T/2T/2T vector magnetic field. For all the magneto-optical measurements, the magnetic field was swept along the *b*-axis of the sample. A 50× objective (NA = 0.55) was used for optical excitation and collection in normal incident geometry. A piezoelectric scanner was used to raster scan the sample for PL and SHG microscopic images with a spatial resolution of ~1 μm. For PL spectroscopy, a 633-nm HeNe laser was used to excite the sample with linear polarization along the *b*-axis, and the PL signal was captured through a spectrograph equipped with a liquid-nitrogen-cooled charge-coupled device. The PL hysteresis loops were recorded by integrating emission from 1.24 eV to 1.38 eV via an avalanche photodiode or the spectrograph. For the SHG measurements, a femtosecond oscillator (InSight X3, Spectra Physics) with ~120 fs pulse duration was used. The laser polarization was along the $\boldsymbol{b}$-axis and the wavelength was 925 nm unless otherwise specified. The SHG intensity was acquired by using a photomultiplier tube in photon-counting mode. The azimuthal SHG polarization patterns were obtained by rotating both the excitation and detection beams with a half-wave plate and setting them in the co- (XX) or cross- (XY) polarizations[17].

**Phase-SHG measurement**

The schematic of the phase-resolved SHG setup is shown in Extended Data Fig. 1 and also detailed in Ref. [45]. A y-cut quartz was placed between the sample and the objective, with the finely tuned angle and distance relative to the sample. For CrSBr, the SHG patterns with linear co- (XX) and cross- (XY) polarizations indicate a significantly stronger XY over XX intensity. The laser polarization was thus aligned to the $\boldsymbol{b}$-axis for all the phase-SHG measurements. The y-cut quartz, exhibiting two-fold symmetry, contributes strong XX SHG signal along its $\boldsymbol{x}$-axis. To conduct the phase-SHG measurements, the $\boldsymbol{x}$-axis of quartz was aligned to the $\boldsymbol{b}$-axis of CrSBr. So, the polarizations of the SHG fields from quartz ($\vec{\boldsymbol{E}}_{quartz}^{2\omega}$) and CrSBr ($\vec{\boldsymbol{E}}_{CrSBr}^{2\omega}$) were orthogonal. The two orthogonal signals passed through a phase-shifting Soleil-Babinet compensator (SBC), whose fast- and slow-axis were aligned with the SHG polarizations of quartz and CrSBr, respectively. The phase difference of SHG between quartz and CrSBr was thus continuously adjusted through the translation of birefringent wedge in SBC. The two SHG signals were then projected to the polarization direction of an analyzer for superposition and interference before being detected by a photomultiplier tube. The phase-SHG interferogram as a function of external phase shift via SBC can thus be obtained. By comparing the phase difference between magnetic polymorphs with time-reversal or spatial-inversion operation, the expected $\pi$ phase shift can be measured. We note that if one needs to obtain the absolute phase value of a specific magnetic state, another independent phase reference would be required for calibration.

**Acknowledgments**

The work at Fudan University was supported by National Key Research and Development Program of China (Grant Nos. 2022YFA1403302, 2019YFA0308404), National Natural Science Foundation of China (Grant Nos. 12034003, 11427902, 91950201, 12004077), Science and Technology Commission of Shanghai Municipality (Grant Nos. 20JC1415900,

21JC1402000, 23JC1400400, 2019SHZDZX01), Program of Shanghai Academic Research Leader (Grant No. 20XD1400300), Shanghai Municipal Education Commission (2021KJKC-03-61), and China National Postdoctoral Program for Innovative Talents (Grant No. BX20200086).

**Author contributions**

S.W.W. conceived and supervised the project. Z.Y. Sun and C.H. performed the experiments with assistance from Z.Y. Sheng, S.W., Z.S.W. and B.K.L. Y.C., Q.M. and Z.L. provided the CrSBr single crystals. Z.Y. Sun, C.H. and S.W.W. analyzed the data and wrote the paper with inputs from W.L., Z.Y., Y.W. and J.S.

**Competing interests**

The authors declare no competing interests.

**Data availability**

All data that support the findings of this study are available from the corresponding authors on reasonable request. Source data are provided with this paper.

**Figures and captions**

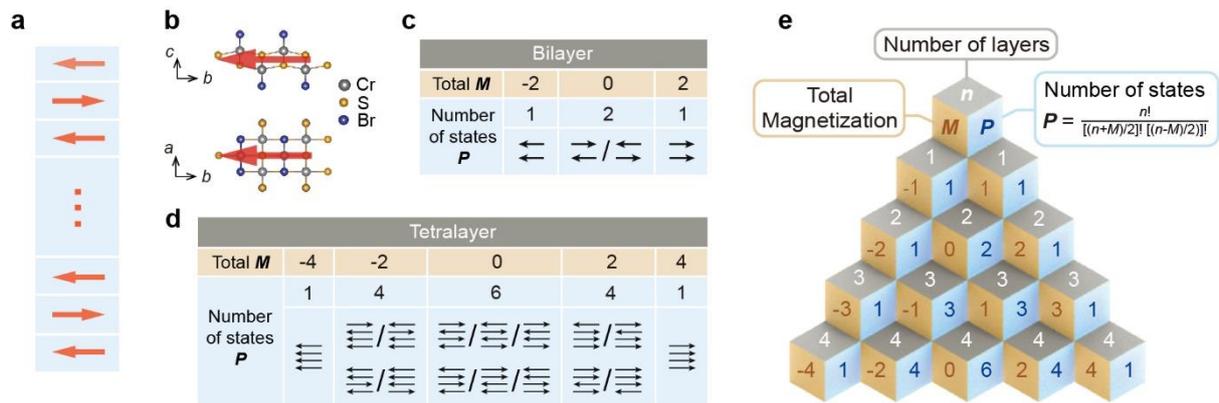

**Fig. 1 | Combinatorial nature in layered antiferromagnets. a**, Schematic of layered AFM order. **b**, Lattice structure of CrSBr monolayer with in-plane ferromagnetism (orange arrows) along **b**-axis. **c,d**, Total magnetization and the corresponding magnetic states in CrSBr bilayer (**c**) and tetralayer (**d**). **e**, *Yang Hui*'s (*Pascal*'s) Triangle for layered antiferromagnets. The number of layers, total magnetization and number of magnetic states are denoted as **n**, **M** and **P**, respectively.

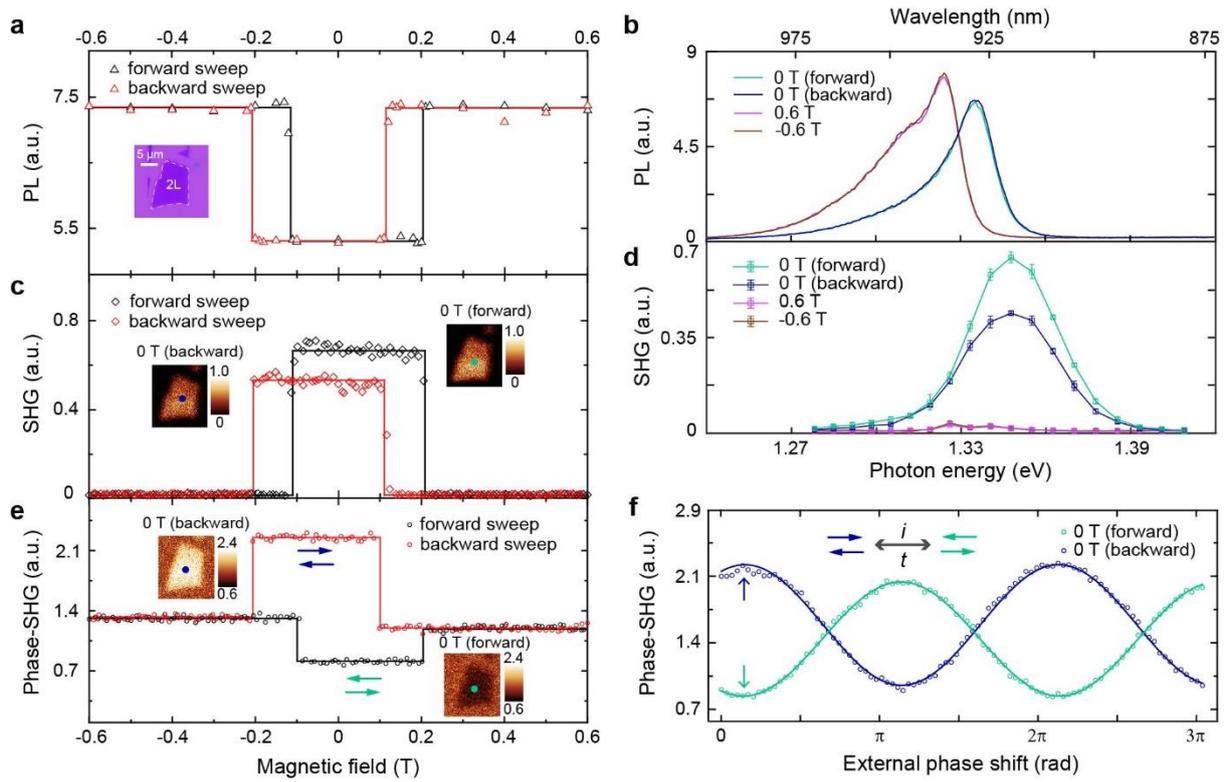

**Fig. 2 | Resolving the layered antiferromagnetism in 2L CrSBr. a**, PL hysteresis loop of 2L CrSBr. The switchings are outlined as step-like solid lines. Inset: optical microscopic image. Scale bar: 5 μm. **b**, PL spectra at 0 T and ±0.6 T. **c**, SHG hysteresis loop. Inset: SHG images at 0 T (backward, left) and 0 T (forward, right). **d**, SHG excitation spectra at 0 T and ±0.6 T. **e**, Phase-resolved SHG hysteresis loop. The external phase shift was set at $\pi/6$. The arrows denote the magnetic structures at two AFM states. Inset: phase-SHG images at 0 T (backward, left) and 0 T (forward, right). **f**, Phase-SHG intensity as a function of external phase shift. The sinusoidal fits are shown as solid lines. The data were taken at the positions marked in (**c**) and (**e**). Inset: The two AFM states form the time-reversal *t* or spatial-inversion *i* counterparts.

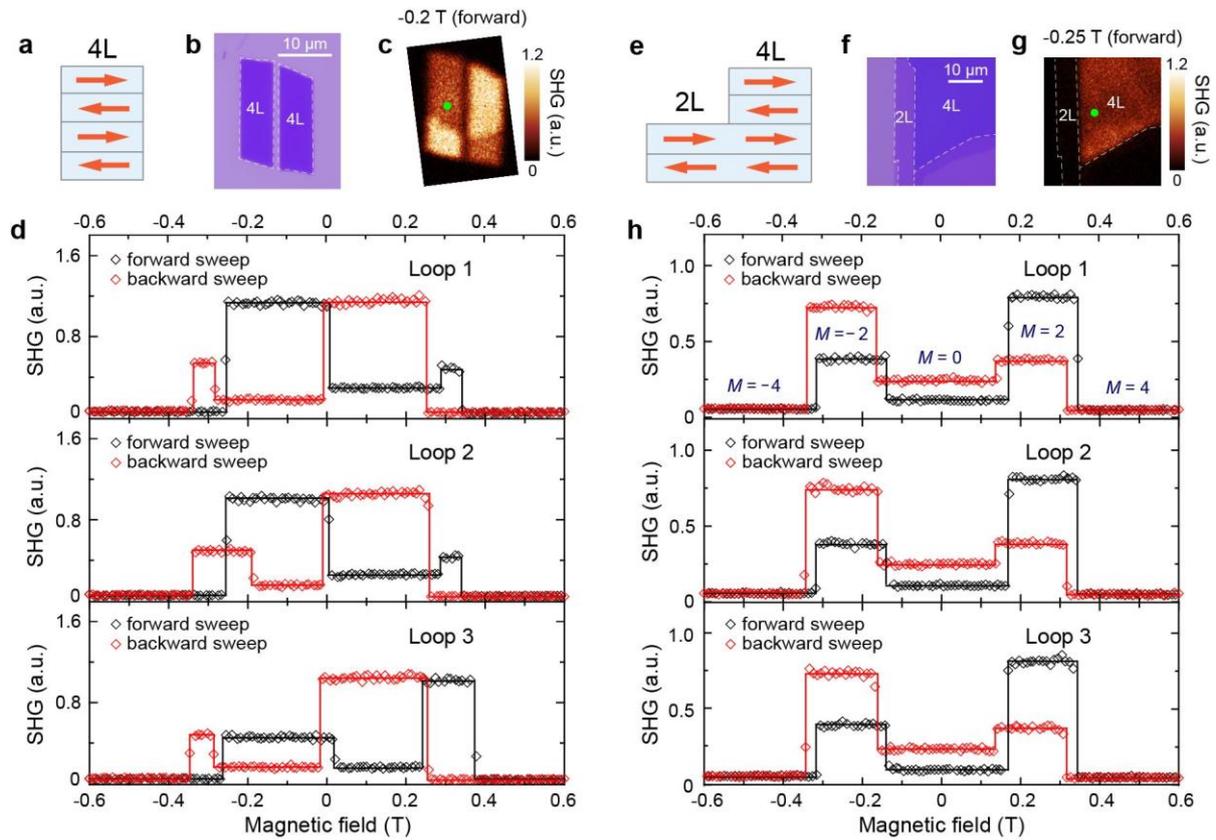

**Fig. 3 | Magneto-SHG hysteresis on 4L CrSBr. a**, Antiferromagnetic ground state of isolated CrSBr tetralayer. **b**, Optical microscopic image of isolated 4L CrSBr. Scale bar: 10 μm. **c**, SHG image of isolated 4L CrSBr at -0.2 T (forward). **d**, Three non-repetitive SHG hysteresis loops of isolated 4L CrSBr obtained at the green dot in (**c**). **e**, Antiferromagnetic ground state of non-isolated CrSBr tetralayer with a laterally extended bilayer. **f**, Optical microscopic image of non-isolated 4L CrSBr. Scale bar: 10 μm. **g**, SHG image of non-isolated 4L CrSBr at -0.25 T (forward). **h**, Three repetitive SHG hysteresis loops of non-isolated 4L CrSBr obtained at the green dot in (**g**).

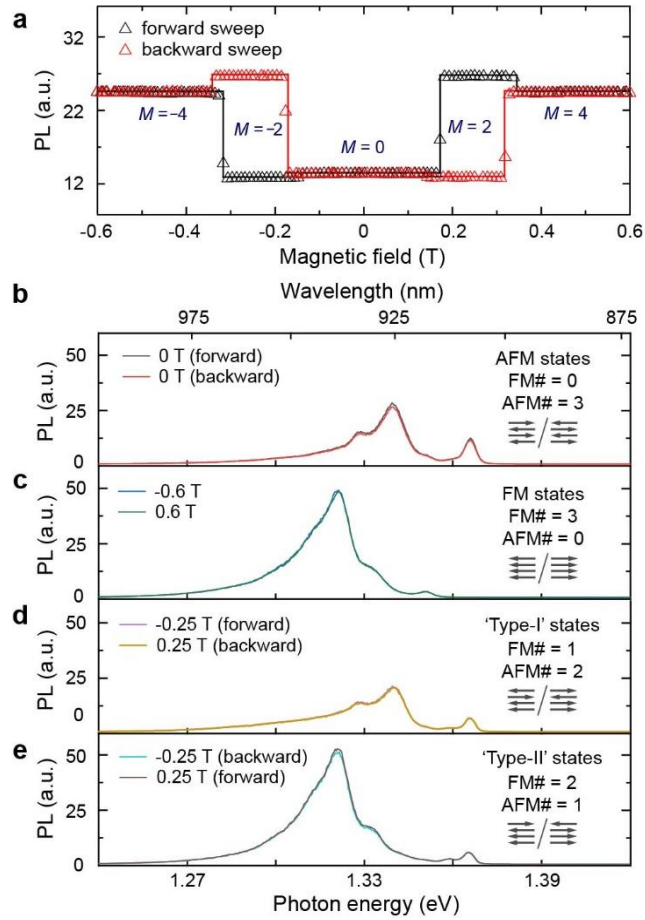

**Fig. 4 | Magneto-PL loop and spectra on the non-isolated 4L CrSBr. a**, PL hysteresis loop. **b-e**, PL spectra at the AFM (**b**), FM (**c**), 'Type-I' (**d**) and 'Type-II' (**e**) states, grouped by the number of FM and AFM interfaces listed in Extended Data Table 1.

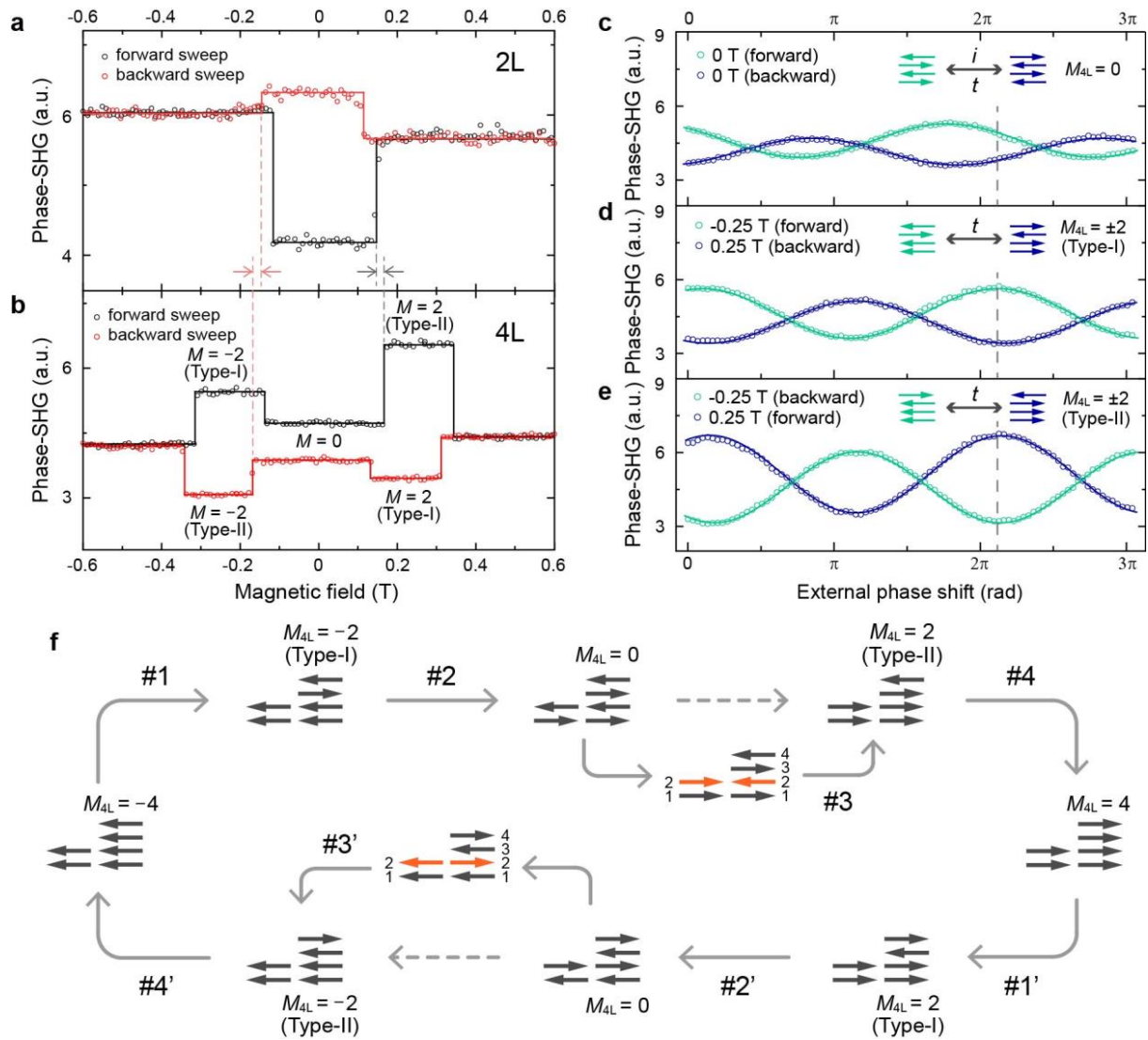

**Fig. 5 | Resolving the magnetic structures and transitions of non-isolated 4L CrSBr and extended 2L. a,b**, Phase-SHG hysteresis loops of non-isolated 4L (**a**) and extended 2L (**b**), with external phase shift set at $11\pi/4$ for 2L and $11\pi/5$ for 4L. Dashed lines highlight the sequential spin-flip transitions initiate from 2L and follow by 4L. **c-e**, Phase-SHG intensity as a function of external phase shift between two AFM (**c**), two 'Type-I' (**d**) and two 'Type-II' (**e**) states. Gray dashed lines show the external phase shift utilized in (**b**). Inset illustrate the magnetic structures of the corresponding states. **f**, Route of the magnetic transitions for 2L and 4L. The spin-flip transitions are divided into several steps as #1-#4 in forward sweep and #1'-#4' in backward sweep.

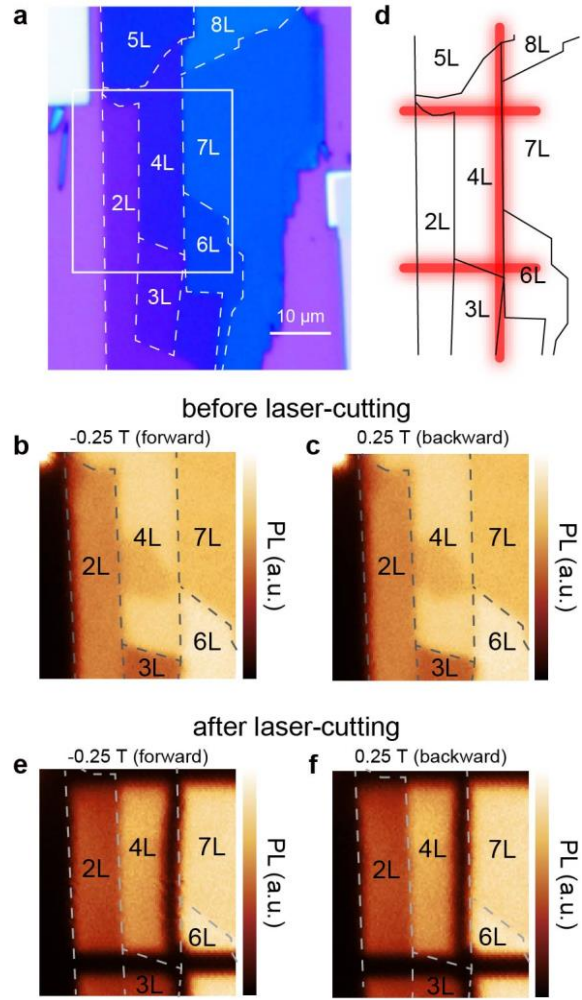

**Fig. 6 | Layer-sharing effect in a few-layer CrSBr. a**, Optical microscopic image of the few-layer CrSBr. The 4L region is adjacent to a 2L, a 3L and thicker layers on the left, bottom and right, respectively. The regions with different thicknesses are delineated by white dashed lines. Scale bar: 10 μm. **b,c,** PL images before laser cutting at -0.25 T (forward) (**b**) and 0.25 T (backward) (**c**). **d,** Schematic of *in-situ* laser cutting on the sample. The cutting paths are highlighted in red lines for clarity. **e,f,** PL images laser cutting at -0.25 T (forward) (**e**) and 0.25 T (backward) (**f**).